\documentclass[12pt,twocolumn]{iopart}
\usepackage{graphicx}
\usepackage{cite}
\begin{document}

\title[Resistivity of BaFe$_{2}$As$_{2}$ under high pressure]{High pressure study of BaFe$_{2}$As$_{2}$ -- role of hydrostaticity and uniaxial stress}

\author{W.~J.~Duncan$^{1}$, O.~P.~Welzel$^{2}$, C.~Harrison$^{1}$, X.~F.~Wang$^{3}$, X.~H.~Chen$^{3}$, F.~M.~Grosche$^{2}$, P.~G.~Niklowitz$^{1}$}
\address{$^1$ Department of Physics, Royal Holloway, University of London, Egham TW20 0EX, UK}
\address{$^2$ Cavendish Laboratory, University of Cambridge, Cambridge CB3 0HE,UK}
\address{$^3$ Hefei National Laboratory for Physical Sciences at Microscale and Department of Physics, University of Science and Technology of China, Hefei, Anhui 230026, P.~R.~China}
\ead{w.j.duncan@rhul.ac.uk; philipp.niklowitz@rhul.ac.uk}
\begin{abstract}

We investigate the evolution of the electrical resistivity of
BaFe$_{2}$As$_{2}$ single crystals with pressure. The samples used
were from the same batch grown from self flux and showed
properties that were highly reproducible. Samples were pressurised
using three different pressure media: pentane-isopentane (in a
piston cylinder cell), Daphne oil (in an alumina anvil cell) and
steatite (in a Bridgman cell). Each pressure medium has its own
intrinsic level of hydrostaticity, which dramatically affects the
phase diagram. An increasing uniaxial pressure component in this
system quickly reduces spin density wave order and favours the
appearance of superconductivity, similar to what is seen in
SrFe$_{2}$As$_{2}$.

\end{abstract}
 \submitted{\JPCM \maketitle

The recent discovery of superconductivity in transition metal
pnictides at temperatures as high as 55\,K has ignited an industry
of research. A number of structure families are being
investigated: the '1-1-1-1 compounds', such as LaOFeAs, the '1-1-1
compounds', such as LiFeAs, the '1-1 compounds', such as FeSe, and
the '1-2-2 compounds', such as SrFe$_2$As$_2$. Among these, the
oxygen free compounds stand out for being comparatively easy to
grow as high quality, homogeneous and stoichiometric, large single
crystals. We concentrate on a key member of the 1-2-2 iron
arsenide family, BaFe$_{2}$As$_{2}$, which, when doped with
potassium~\cite{hu_08a, rot08a, sas08a, jee08a}, has the highest
superconducting transition temperature, $T_{c}$, of all the
oxygen-free iron arsenide compounds.

The 1-2-2 compounds CaFe$_2$As$_2$~\cite{ni_08a},
SrFe$_2$As$_2$~\cite{pfi80a} and
BaFe$_2$As$_2$~\cite{rot08b,hua08a} undergo a magnetostructural
transition into a spin density wave state on cooling. Their low
temperature state can be modified effectively by substituting iron
with a number of other transition metal elements, by substituting
the alkaline earth element with potassium or by substituting
arsenic with phosphorus. All of these approaches can be used to
suppress the magnetostructural order of the parent compounds,
giving rise -- in most cases -- to superconductivity at elevated
temperatures of the order of 20-40\,K. The resulting phase diagram
is similar to that of numerous heavy fermion
systems~\cite{mat98a}, organic superconductors~\cite{lef00a} and,
more recently, an alkali metal fulleride compound~\cite{tak09a}.
This generality points to a fundamental connection between
magnetism and superconductivity in these strongly correlated
electron systems.

Alternatively, the low temperature phase diagram of the 1-2-2
compounds can be investigated by applying pressure. Several high
pressure studies have been published within a short time,
beginning with the discovery of pressure-induced superconductivity
in CaFe$_2$As$_2$~\cite{tor08a}. Usually, pressure tuning has
important advantages. It does not vary the disorder level, it can
be applied with great precision, allowing access to the closest
proximity of a quantum phase transition, and it is highly
reproducible. The pressure studies on the 1-2-2 compounds, by
contrast, have led to a bewildering array of confusing and
apparently contradictory results. At first, it seems quite
straightforward to explain these discrepancies by the difference
in sample quality; in the previous studies either polycrystals or
single crystals were used and the residual resistivity ratio (RRR)
was found to vary between 1.4 and 10. However, pressure results in
the 1-2-2 iron arsenide compounds appear to scatter more wildly
than the results of chemical substitution studies, suggesting an
additional factor causing the discrepancies. This could be the
difference in hydrostaticity caused by the usage of different
pressure media in different pressure cells. The effect of the
level of hydrostaticity has now been studied in both
CaFe$_{2}$As$_{2}$ \cite{chu09b,par08a,tor08a} and
SrFe$_{2}$As$_{2}$~\cite{kot09a} but a comprehensive study for
BaFe$_2$As$_2$ is still lacking.





At room temperature and ambient pressure, BaFe$_2$As$_2$ has the
tetragonal (I4/mmm) ThCr$_{2}$Si$_{2}$ structure~\cite{pfi80a}.
Below 135\,K, it undergoes a magnetostructural transition to an
orthorhombic spin density wave (SDW) phase~\cite{rot08b}.
In this phase, the Fe atoms acquire magnetic moments of 0.87(3)
${\mu}_{B}$ with an ordering wavevector {\bf Q} =
(101)~\cite{hua08a,kof09a}. Band structure calculations suggest
that the spin density wave instability can be attributed to
nesting between electron and hole Fermi surface sheets
\cite{kim09a, sin08a, she08a ,zha08a}. With increasing pressure or
doping~\cite{kim09a}, this nesting degrades, leading to a gradual
suppression of the spin density wave
order. 


\begin{table}
\begin{center}
\begin{tabular}{|l|r|c|c|c|c|}

\hline
Pressure medium & Technique & $T_{c,max}$ & $p_{max}$ & $dT_{SDW}/dp$ & Ref. \\
&         &        (K)        &   (kbar)       & $~\rm{(K kbar^{-1})}$ & \\
\hline
Daphne Oil 7373     & DAC   &   29      &   40      &   NA  &  \cite{ali09a}      \\
Fluorinert 70/77 1:1    & CAP   &   30      &   $\sim$ 35   &   -1.35   &   \cite{fuk08a}     \\
Fluorinert 70/77 1:1    & BC    &   30      &    53   &   -2.2    &   \cite{col09a} \\
Steatite        & BC    &   35.4      &    15       &   -2.43   &   \cite{man09a}      \\
Glycerin        & CAP   &   -      &   (80)      &   -0.7    &   \cite{mat09b}  \\
Daphne Oil 7373     & PCC   &   -      &   (24)      &   -0.76   &   \cite{ahi09a}       \\
Pentane-Isopentane 1:1  & PCC   &   -      &   (30.7)      &   -0.84   &   this study       \\
Daphne Oil 7373     & AAC   &   $\geq 24.5$        &   $\geq 55$      &   -1.09   &   this study       \\
Steatite        & BC    &   32.5        &   10.6      &   -  &   this study       \\

\hline
\end{tabular}
\end{center}
\caption{Pressure studies in BaFe$_2$As$_2$. The table summarises
work carried out using different pressure media and techniques
(DAC: diamond anvil cell, CAP: cubic anvil press, BC: Bridgman
cell, PCC: piston-cylinder cell, AAC: alumina anvil cell). $T_c$
gives the maximum observed transition temperature in the
superconducting dome. This is the onset of a drop in
magnetisation~\cite{ali09a} or the onset of a resistance drop
(other work). Zero resistance is only reported in two of the
studies~\cite{col09a,man09a}. $p_{max}$ gives the pressure at
which $T_c$ is maximal. (If no superconductivity has been
observed, $p_{max}$ gives the maximum pressure of the experiment.)
$d T_{SDW}/d p$ denotes the drop of the spin density wave
transition temperature in the zero-pressure limit.}
\label{pressureSummary}

\end{table}

A number of high pressure studies have been carried out on
BaFe$_2$As$_2$~\cite{fuk08a,ali09a,col09a,man09a,ahi09a,mat09b}
(Table~\ref{pressureSummary}). In all cases, pressure application
suppresses the magnetostructural transition to lower temperatures.
However, the rate of decrease of the spin density wave transition
temperature, $T_{SDW}$ with pressure differs greatly between these
studies. Additionally, the extent to which indications for
superconductivity are observed varies strongly. Whereas one study
reports a diamagnetic signal indicating superconductivity in a
large volume fraction of the sample~\cite{ali09a}, other studies
show incompleteness~\cite{fuk08a} or even absence of
superconducting transitions in the
resistivity~\cite{ahi09a,mat09b}. In the studies showing signs of
superconductivity the maximum transition temperature $T_{c,max}$
and the pressure under which it occurs, $p_{max}$, vary
considerably, as does the pressure range, across which
superconductivity has been observed. So far, it cannot be said,
whether the observed variations are mainly due to differences in
the sample quality or rather due to differences in the employed
pressure media.

To separate these issues, we present and compare high pressure
data obtained from the {\it same} batch of high-quality single
crystals of BaFe$_2$As$_2$ subject to three {\it different}
pressure media: (i) pentane-isopentane (used in a piston cylinder
cell up to 3 GPa), (ii) Daphne oil 7373 (used in an opposed alumna
anvil cell up to 6 GPa) and (iii) steatite (used in a Bridgman
cell up to 7 GPa). We expected nearly ideal hydrostatic conditions
for method (i), and progressive deviation from hydrostaticity with
methods (ii) and (iii). Due to the used pressure cell geometries,
it is expected that deviations from hydrostaticity include
significant uniaxial pressure components. Our results suggest that
even very moderate amounts of uniaxial stress induce at least
filamentary superconductivity in BaFe$_2$As$_2$. Stronger uniaxial
stress fundamentally changes the phase diagram, leading to a fast
suppression of the orthorhombic spin density wave phase.




The samples were grown using a self flux method, which yielded
single crystals that were typically 50 ${\mu}$m thick and weighed
several mg. All measurements were conducted using a Quantum Design
Physical Properties Measurement System (PPMS). The resistivity was
measured using an AC four point technique with the current in the
a-b plane and the magnetic field parallel to the c-axis. Contacts
were made by spot welding 25 ${\mu}$m gold wire onto the sample,
except in the case of the Bridgman cell measurements, in which the
contacts consisted of 25 ${\mu}$m platinum wire pressed onto the
sample. The pressure was determined from the superconducting
transition temperature of a lead sample in the alumina anvil and
Bridgman cells and of a tin sample in the piston cylinder cell.
The pressure inhomogeneity was estimated from the width of the
superconducting transition produced by the lead or tin sample. The
BaFe$_2$As$_2$ crystals were characterized by resistivity and heat
capacity measurements at ambient pressure. The samples showed
properties similar to samples reported in the
literature~\cite{rot08a, chu09a,
 rot08a} including a spin density wave transition temperature
$T_{SDW}=131$\,K.

\begin{figure}[]
\begin{center}
      \includegraphics[width=0.7\columnwidth]{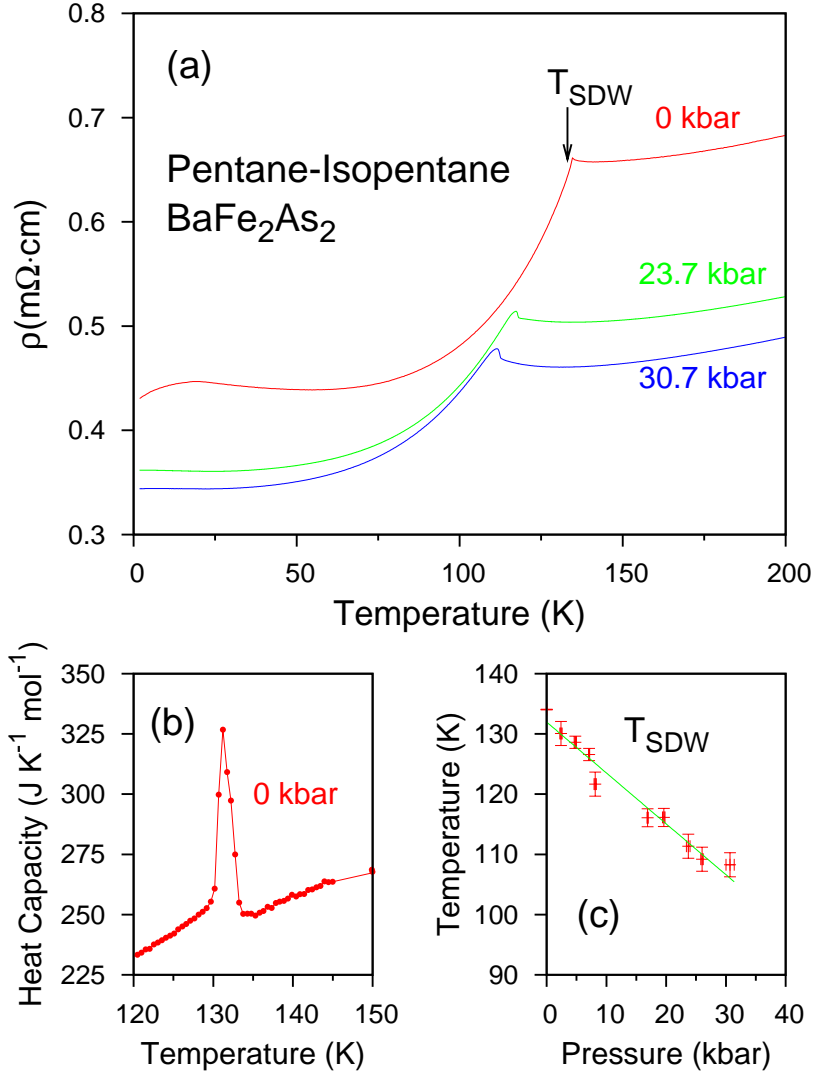}
\end{center}
\caption{Measurements using pentane-isopentane as the pressure
medium in a piston-cylinder cell. The magnetostructural transition
($T_{SDW}$) of BaFe$_{2}$As$_{2}$ is clearly visible (a) in the
resistivity and (b) in the heat capacity at zero pressure. Under
pressure (see (a) and (c)) the magnetostructural transition is
suppressed at a rate of $\sim$ -0.84 K kbar$^{-1}$.}
\label{cylinder}
\end{figure}

The first set of measurements was conducted in a piston cylinder
cell (Figure \ref{cylinder}), using a 1:1 mixture of pentane and
isopentane as pressure medium. According to the width of the
superconducting transition of the tin manometer, these
measurements produced the most hydrostatic conditions of the three
pressure methods employed in this study (Figure \ref{inhomo}). The
resistivity was measured to a maximum pressure of 30.7 kbar. The
magnetostructural transition, determined from the maximum of
$d\rho/dT$, is slowly suppressed at a rate of approximately
$-0.84$\,K\,kbar$^{-1}$. At the maximum pressure, the spin density
wave transition is still clearly visible with no signs of
broadening, which indicates that the pressure remains hydrostatic.
No anomaly suggestive of superconductivity was observed at low
temperatures. At ambient pressure there is a broad maximum in the
resistivity around 19\,K, which disappears above 8\,kbar, similar
to what is seen by Matsubayashi et al~\cite{mat09b}. The origin of
this hump in the resistivity trace is unclear. It is not
associated with any signature in the heat capacity.

\begin{figure}[]
\begin{center}
      \includegraphics[width=0.7\columnwidth]{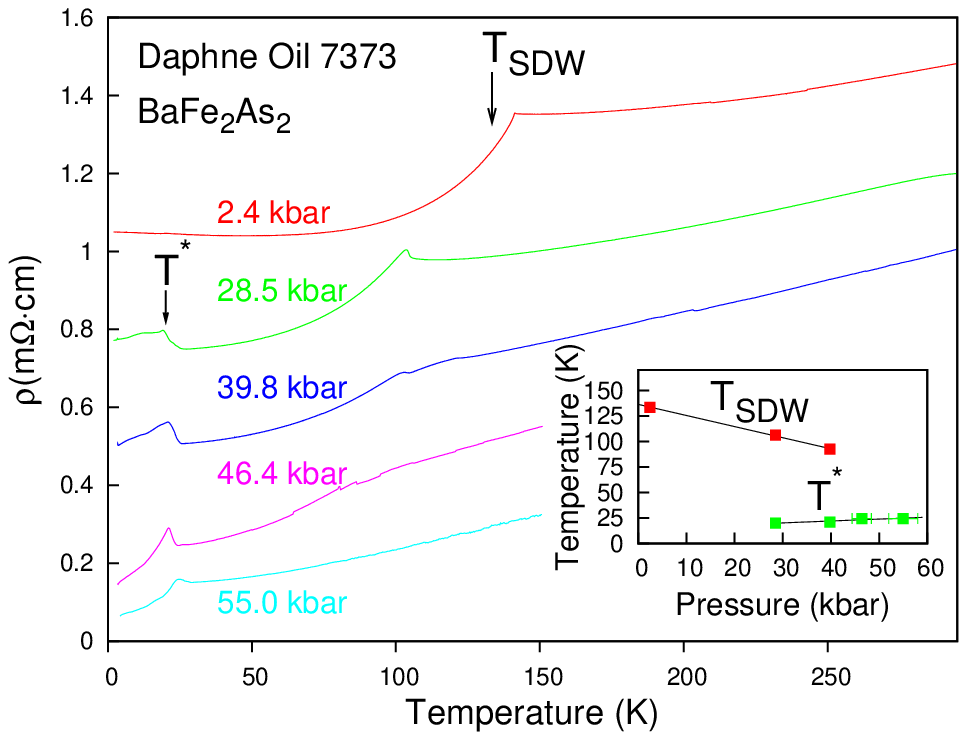}
\end{center}
\caption{Resistivity measurements of BaFe$_{2}$As$_{2}$ using
Daphne oil 7373 as the pressure medium in an opposed alumina anvil
cell. Curves are shifted for clarity. In the inset, the pressure
evolution of the superconducting onset (T$^{*}$) and of the
magnetostructural (T$_{SDW}$) transition is shown. T$_{SDW}$ is
suppressed at a rate of $\sim$ -1.09\,K\,kbar$^{-1}$.}
\label{fig3}
\end{figure}

The second set of measurements was conducted in an alumina anvil
cell, in which the sample was aligned with the $c$-axis
perpendicular to the anvil flats. The sample space was filled with
Daphne oil 7373 as pressure medium. This appears to offer slightly
less hydrostatic conditions than pentane-isopentane, possibly due
to its increased viscosity (Figure \ref{inhomo}). In this case
(Figure \ref{fig3}) the magnetostructural transition is initially
suppressed at a slightly higher rate of $-1.09$\,K\,kbar$^{-1}$,
compared to the piston cylinder cell and is no longer visible at
46.4\,kbar. At low temperatures, an anomaly (labelled $T^{*}$)
appears at 28.5\,kbar, where the resistivity has a maximum near
20\,K. As the pressure is increased further, this feature grows
into a sharp drop, which is largely pressure independent.
Behaviour similar to the one at $T^{*}$ has been previously
associated with filamentary superconductivity in previous studies
(e.g., Ref.~\cite{fuk08a}). It is also interesting to note that
the pressure regimes where $T^*$ and $T_{SDW}$ are seen, overlap.

\begin{figure}[]
\begin{center}

      \includegraphics[width=0.65\columnwidth]{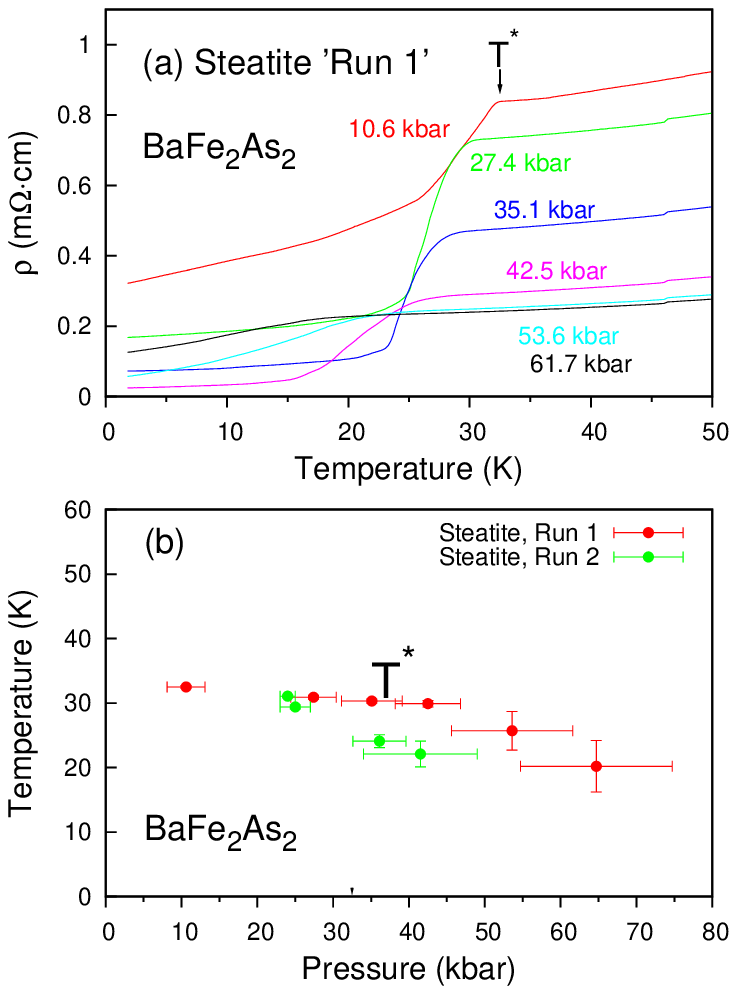}
\end{center}
\caption{Resistivity measurements of BaFe$_{2}$As$_{2}$ using
steatite as the pressure medium in a Bridgman cell. A
representative data set ('Run 1') is shown in (a). The pressure
evolution of the onset of the superconducting transition T$^{*}$
is shown in (b).} \label{fig4}
\end{figure}

\begin{figure}[]
\begin{center}
      \includegraphics[width=0.7\columnwidth]{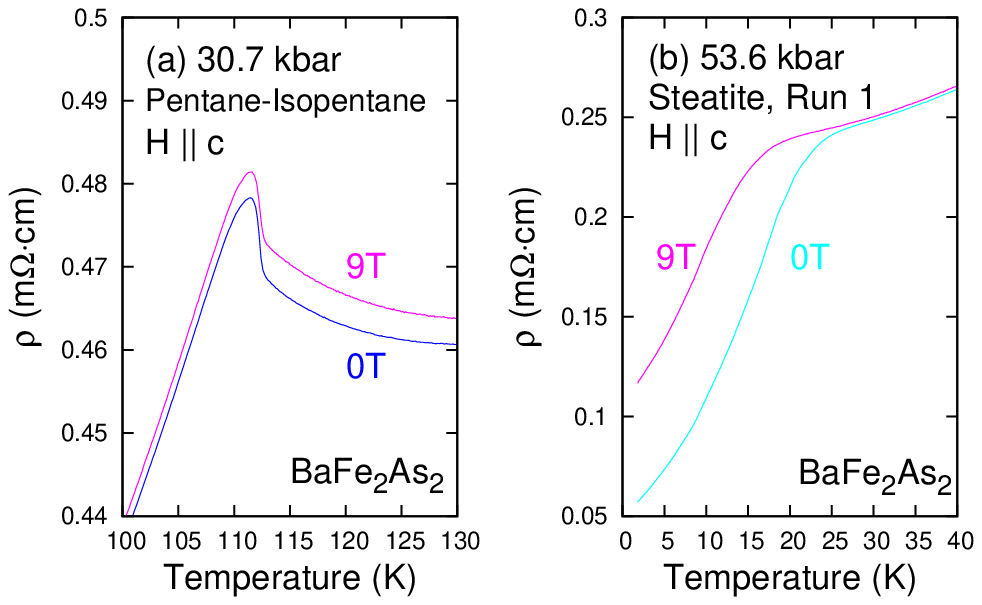}
\end{center}
\caption{Effect of a magnetic field on the observed transitions.
(a) The magnetostructural transition (T$_{SDW}$) in the
resistivity measurement at 30.7\,kbar using pentane-isopentane as
the pressure medium in a piston cylinder cell. This transition
does not show any significant field dependence (b) The
superconducting transition (T$^{*}$) in the resistivity
measurement at 53.6\,kbar using steatite as the pressure medium in
a Bridgman cell. This transition has a clear field dependence.}
\label{fig5}
\end{figure}

The third set of pressure experiments was carried out using
steatite as the pressure medium in a Bridgman cell, again with the
c-axis of the sample normal to the anvil flats. This setup has
been expected to provide the largest uniaxial pressure component.
Already at the lowest pressure measured in this cell, 10.6 kbar
(Figure~\ref{fig4}a), there is no sign of the magnetostructural
transition. Because the jump in the resistivity at low
temperatures is similar to the anomaly seen in the alumina anvil
cell at 55 kbar, we also label this transition $T^{*}$. In this
case $T^*$ starts at a slightly higher temperature of ${\sim}
32$\,K, which is comparable to the superconducting onset in other
studies~\cite{fuk08a,ali09a, kim09a,
 col09a}. With increasing pressure, the resistivity curves look similar to what is observed by
Fukazawa et al~\cite{fuk08a}. $T^{*}$ was found to be clearly
field dependent: a magnetic field of 9\,T applied at 53.6 kbar
suppressed the transition by $0.80$\,K\,T$^{-1}$
(Figure~\ref{fig5}b). These observations and the fact that $T^{*}$
is weakly pressure dependent suggest that $T^*$ represents the
onset temperature of partial or filamentary superconductivity. Our
findings were reproduced in a second sample.

\begin{figure}[]
\begin{center}
      \includegraphics[width=0.7\columnwidth]{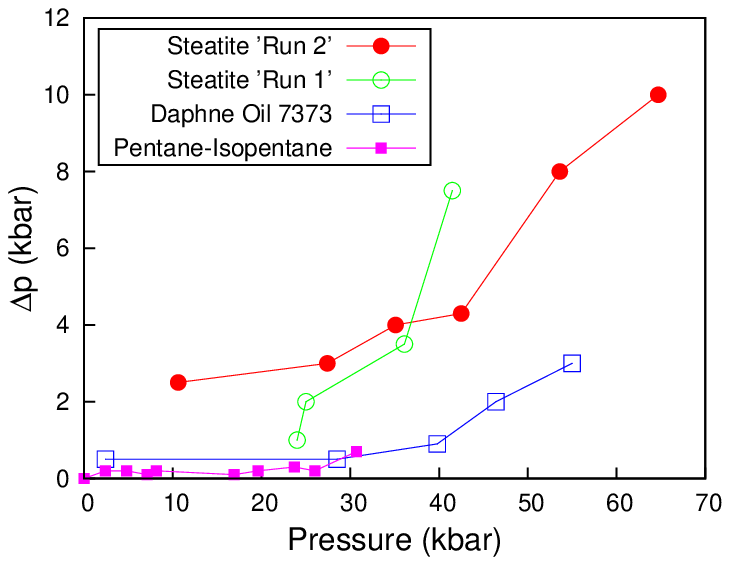}
\end{center}
\caption{The pressure inhomogeneity ($\Delta p$) for each of the
  different pressure media as a function of pressure, as calculated
  from the width of the superconducting transition of the pressure
  gauge. The pressure media in order of highest hydrostaticity are
  pentane-isopetane, Daphne oil 7373, and steatite.}
\label{inhomo}
\end{figure}

The pressure-temperature phase diagrams obtained by the three
different high pressure methods (Figures \ref{cylinder}c,
\ref{fig3} (inset) and \ref{fig4}b) are dramatically different: no
superconductivity is observed at all up to 30\,kbar in the sample
floating in pentane-isopentane, whereas the onset of at least
filamentary superconductivity appears already at 10\,kbar in a
sample embedded in steatite. These results demonstrate that the
precise pressure conditions strongly influence the high-pressure
properties of BaFe$_2$As$_2$. Since the minimal pressure for the
onset of superconductivity or the critical pressure for the
suppression of spin density wave order shifts by several tens of
kbar in different pressure setups, simple pressure inhomogeneity
(pressure gradients across the sample) does not explain the
observed differences. Instead, non-hydrostaticity in opposed-anvil
setups is expected to arise in form of considerable uniaxial
components leading to uniaxial stress on the sample. Therefore,
our measurements show that increased uniaxial pressure components
along the $c$-axis suppress the spin density wave order at a
faster rate. Uniaxial pressure components also favour the
appearance of a regime of at least filamentary superconductivity,
which extends to lower pressures, when they become stronger.

A comparison of our results with other high-pressure data on
BaFe$_2$As$_2$ (Table~\ref{pressureSummary}) shows that the
variations in previously observed phase diagrams can be explained
by the differences in the pressure conditions alone. This follows
from our result that we can reproduce a similar range of different
phase diagrams using samples from the same batch with the same
sample quality. It still remains to be seen, whether pressure
induced {\it bulk} superconductivity is intrinsic to
BaFe$_2$As$_2$. So far, one study reported evidence for bulk
superconductivity in BaFe$_2$As$_2$~\cite{ali09a}. However, there
are studies involving very high (Ref.~\cite{mat09a}) and quite low
(our measurements in steatite) levels of hydrostaticity, in which
bulk superconductivity is absent up to high pressures.

The dependence of the pressure-temperature phase diagram of
BaFe$_2$As$_2$ on the level of hydrostaticity is reminiscent of
what has been reported in the case of
SrFe$_2$As$_2$~\cite{kot09a}. There, the phase diagram is
influenced in a qualitatively similar way by uniaxial pressure
components, although the effect is more dramatic in the case of
BaFe$_2$As$_2$. Similarly, in CaFe$_2$As$_2$, superconductivity
near the structural transition only appears when there is
sufficient pressure inhomogeneity. The resulting shear stress
gives rise to a metastable phase and superconductivity is absent
when a helium pressure medium is used~\cite{yu_09a}. An exception
is the spin density wave order of CaFe$_2$As$_2$, which is not
more strongly suppressed by less hydrostatic~\cite{tor08a} or even
uniaxial pressure conditions~\cite{tor09a}. The comparison of our
and previous results on 1-2-2 compounds shows that sensitivity to
the precise pressure conditions is a generic phenomenon of this
material class.

The stronger suppression of spin density wave order in
BaFe$_2$As$_2$ by a uniaxial pressure component compared to
hydrostatic pressure might be a consequence of the effects of
uniaxial stress on the Fermi surface. Stronger reduction of the
$c/a$ ratio will tend to increase interplane hopping and warping
of originally cylinder-like Fermi surface sheets. As a consequence
nesting will be reduced, which decreases the possibility for spin
density wave order to form. For testing this interpretation it is
best to focus on studies, which provide the most direct link
between lattice and electronic properties. This includes studies
using pressure tuning or charge-neutral chemical substitution but
excludes studies involving electron or hole doping. Examples for
the suppression of spin density wave order being accompanied by a
reduction of the $c/a$ ratio are tuning BaFe$_2$As$_2$ by
pressure~\cite{kim09a} or substitution of As by P~\cite{jia09a} or
tuning SrFe$_2$As$_2$ by Ru substitution~\cite{sch09a}.

The ways to optimise any superconducting transition temperature
($T_c$) might be independent from the best way to suppress spin
density wave order. For optimising $T_c$ several lattice
parameters have been proposed as key quantities: the $c/a$ ratio
in connection with SrFe$_2$As$_2$~\cite{kot09a}, the pnictogen
height in connection with NdFeAsO and LaFePO (which is also
supposed to influence the symmetry of the order
parameter)~\cite{kur09a}, or the As-Fe-As bond angles in
connection with CeFeAsO$_{1-x}$F$_x$ and BaFe$_2$As$_2$ (which
should approach the value for an ideal
tetrahedron)~\cite{kim09a,zha08b}. Our and previous data on
BaFe$_2$As$_2$ (Table~\ref{pressureSummary}) suggests that the
highest values for $T_c$ are found at the lowest pressures. This
implies that an increase of the $c/a$ ratio and of the volume
within the non-magnetic regime helps raising $T_c$. However,
before a final answer can be expected, the intrinsic nature of
superconductivity in BaFe$_2$As$_2$ has to be better established.




In summary, our investigation in three different pressure
environments demonstrates that the pressure-temperature phase
diagram of BaFe$_2$As$_2$ is extremely sensitive to the precise
pressure conditions and, in particular, to the level of resulting
uniaxial stress. Reducing the $c/a$ ratio of a magnetically
ordered FeAs compound appears to suppress spin density wave order
and favour superconductivity.



\section{References}

\bibliographystyle{unsrt}

\end{document}